\documentstyle[aps,pre]{revtex}
\draft
\input{psfig}
\begin{document}
\twocolumn[\hsize\textwidth\columnwidth\hsize\csname@twocolumnfalse\endcsname
\title{The dynamics of  vortex lines
in the three-dimensional
complex Ginzburg-Landau equation: instability, stretching,  
entanglement, and helices}
\author{I. S. Aranson$^{1,2}$,   A. R. Bishop$^3$ and L. Kramer $^4$ } 
\address{
$^1$ Argonne National Laboratory,
9700 South  Cass Avenue, Argonne, IL 60439\\
$^2$ Department of Physics,
Bar Ilan University, Ramat Gan 52900, Israel\\
$^3$ Theoretical Division and Center for Nonlinear Studies, \\
Los Alamos National Laboratory, Los Alamos, NM 87545 \\
$^4$ Department of Physics, University of Bayreuth, Bayreuth   
95440,   Germany}
\date{\today}
\maketitle
\begin{abstract}
The dynamics of curved vortex filaments
is studied analytically and numerically in the framework of
a three-dimensional
complex Ginzburg-Landau equation (CGLE).
It is shown
that a straight vortex line is unstable with respect to
spontaneous stretching and bending
in a substantial range
of parameters of the CGLE, resulting in formation of persistent
entangled vortex configurations.
The boundary of the three-dimensional instability in parameter  
space is determined.
Near the stability boundary, 
the supercritical saturation of the instability is found, 
resulting in the formation of stable helicoidal vortices.

\end{abstract}
\vskip1pc]
\narrowtext
\section{Introduction}
Analysis of "universal" models plays a central role in
contemporary nonlinear dynamics. Such  models, as with 
Ginzburg-Landau and Swift-Hohenberg equations,
allow for quantitative description of arbitrary nonlinear
system at the threshold of specific instabilities.
The complex Ginzburg-Landau equation (CGLE),
derived some 20 years ago by Newell
\cite{newell} and Kuramoto \cite{kuramoto}
has become a paradigm model
for a qualitative description of weakly-nonlinear oscillatory
media (see for review \cite{cross}). Under appropriate scaling of
the physical variables, the equation assumes the universal form
\begin{equation}
\partial_t A=A-(1+ic)|A|^2A+(1+ib) \Delta A,
\label{cgle}
\end{equation}
where $A$ is the complex amplitude, $b$ and $c$ are real parameters, 
and $\Delta=\partial_x^2+\partial_y^2+\partial_z^2$ is a  
three-dimensional
Laplace operator. The parameter $b$ is the  ratio of dispersion  to
diffusion and $c$ is the ratio of conservative to dissipative  
nonlinearity.
Although the equation is formally valid only at the threshold of
a supercritical Hopf bifurcation, it has been found that the CGLE
often reproduces qualitatively
correct phenomenology over a much wider range of the parameters. As a  
results, the predictions drawn from the analysis of the CGLE (mostly  
in one and
two spatial dimensions, see e.g. \cite{aakw,akw,akw2,chate}) were
recently successfully confirmed by experiments in optical and
chemical systems \cite{arrechi,flessel}.
Moreover, some results obtained from the CGLE (for example,  
symmetry breaking
of spiral pairs) was instructive for interpretation of
experiments in far more complicated systems of
chemical waves \cite{perez} and colonies of amoebae
\cite{alt,gold}

In three dimensions the point singularity at the center of the  
spiral becomes a 
line singularity, known as a scroll, or vortex filament. The  
filaments can be
open (scrolls), closed (vortex loops and rings), knotted or even  
interlinked
or entangled. Scroll vortices had been observed in
 slime mold \cite{siegert}, heart tissues \cite{gray}, gel-immobilized 
 Belousov-Zhabotinskii (BZ) reaction \cite{vilson}.
Long-lived entangled vortex patterns in three-dimensional
BZ reactions were observed
recently by the group of Winfree using advanced optical tomography technique
\cite{winfree}. Complicated vortex configurations have also been 
observed in
numerical simulations of
reaction-diffusion
equations \cite{winfree1,bict,karma}

Theoretical investigation of scroll vortices in reaction-diffusion  
systems was
initiated by Keener \cite{keener}, who derived the equation of  
motion for
the filament axis. In particular, he obtained that the collapse rate is
proportional
to the local curvature of the filament, leading to a collapse of  
a vortex ring
in  a finite time. The existence of non-vanishing vortex  
configurations and
expansion of vortex loops,
observed also in numerical simulations of  reaction-diffusion
equations, was associated
with "negative line tension" of the vortex
filament
\cite{bict}.

Recently, the dynamics of {\it three-dimensional}
(3D) vortex lines  in the CGLE  has
attracted
substantial attention \cite{frisch,gog,ab}.
As a definition  of a vortex line we accept  a  line singularity of the 
phase of a complex function $A$.
Gabbay Ott and Guzdar \cite{gog}
applied a generalisation of Keener's method for
a scroll vortex in reaction-diffusion systems \cite{keener}. They  
derived
that the ring of a radius $R$ collapses in  finite time
according to the following evolution law
 \begin{equation}
\frac{dR}{dt}=-\frac{1+b^2}{R}.
\label{ott}
\end{equation}
In addition, there is no (at least, in the first order in $1/R$)
overall drift of the vortex ring in the direction perpendicular
to the collapse motion. The collapse rate (called often "line
tension") $\nu=1+b^2$ appears to be
in a reasonable agreement with the simulations \cite{gog}.
This result
generalises  Keener's ansatz by including the curvature-induced
shift of the filament's wavenumber.
Thereby, as follows from Eq. (\ref{ott}),  vortex loops initially  
existing
in the system will always shrink  (if, of course, there is no bulk  
instability of the waves emitted by the vortex filament),
and under no condition can
the vortex loop  expand.

In this article we show that under very general conditions,  and in
an extensive part of the parameter space vortex filaments {\it  
expand and
bend} spontaneously
and result in persistent vortex configurations even if there is no  
bulk instability of the emitted waves and the spiral wave is stable in a 
two-dimensional
system.
A preliminary account of this work had been  published in Ref.  
\cite{ab}.
We have shown  that  vortex loops may expand for any value of $b$  
above some
critical value   $b_c(c) $.
The critical value $b_c(c)$ can be relatively small for not too  
large $c$.
For example, our analysis predicts that $b_c \approx 2 $ for $c=0$. 
We prove that for  $b> b_c$ the equation  (\ref{ott})
is not valid, because  formally higher-order, but in fact singular  
corrections,
omitted in Eq. (\ref{ott}), cause severe instability of the filament  
and persistent  stretching. This instability is a three-dimensional 
manifestation of the two-dimensional core instability of spiral waves
(called in Ref. \cite{akw2} {\it acceleration instability}). Its
origin can be traced back to the breakdown of the Galilean  
invariance of the CGLE for
any $\epsilon=1/b\ne0$, causing spontaneous acceleration of
the spiral waves \cite{akw2}. Whereas
in 2D the instability is relatively weak,
the situation is different in three dimensions.
Using combined analytical and numerical methods, we have proven that 
three-dimensional instability of the vortex filaments   persists  
far beyond the 
core instability limit of two-dimensional spiral wave and  
typically has a much
higher growth rate. It ceases to exist
only when the core modes becomes strongly damped. This instability
is not driven by "negative line tension'. It develops from  
a nontrivial response
of the filament core to bending, which results in additional  
"acceleration"
terms in
the equation of filament motion.
As we will show, the bending of the
filament greatly enhances the instability, and may  result in  
formation, after some
transient, of a highly entangled and dense vortex configuration.
The "high dispersion limit"  $b \gg 1$ is readily  fullfilled for  
many physical and
chemical systems (actually $b$ larger than about $2$ is enough to  
have the effect).
For example, in the context of nonlinear optics,
where the CGLE can be  derived from the
Maxwell-Bloch equation in the good cavity
limit \cite{lmn},
this parameter is very small: $\epsilon=1/b \sim 10^{-4}-10^{-3}$.  
For an oscillating chemical reaction the diffusion rates  of various 
reacting components
can be varied over a wide range by adding extra chemicals.

The structure of the article is following. In Sec. II we present an  
analysis
of the filament motion in the "high dispersion" limit   
$\epsilon=1/b \ll 1$. This  allows us to
prove instability of a straight vortex filament with respect to
bending and stretching on the basis of a
computer-assisted analytical procedure.
In Sec. III we present the
results of numerical simulations of the three-dimensional CGLE.
We discuss the properties of a spatio-temporal intermittency
which we have found in our simulations.
In Sec. IV we consider the
formal stability analysis of a straight vortex filament with respect to 
periodic perturbations along the filament axes.
Using a numerical matching-shooting
algorithm, we have calculated the spectrum of eigenvalues and  
determined the
stability limit of a three-dimensional vortex line in the $b,c$ plane.
In Sec. V we present a weakly-nonlinear analysis for the vortex filament
near the three-dimensional stability limit showing the existence of  
traveling
helix solutions. After some concuding remarks in Sec. VI we present
in the Appendix an analytical derivation of
the equation of motion of the vortex line and the stability boundary 
in the limit of the perturbed
nonlinear Schr\"odinger equation (NSE).

\section{Instability of weakly-curved filaments in a high
dispersion limit}

\subsection{Derivation of the equation of motion}

As a test for instability, we consider the dynamics of a weakly curved  
vortex filament in the high-dispersion limit $b \gg 1 $.
We apply the generalisation of the method of Ref. \cite
{akw2} for the case of 3D vortices, and  make perturbations near
the 2D spiral wave solution to the
CGLE. For convenience, we redefine
${\bf r} \to {\bf r}/\sqrt{b}$. Then Eq. (\ref{cgle})  assumes the
form
\begin{equation}
\partial_t A=A-(1+ic)|A|^2A+(\epsilon+i) \Delta A.
\label{cgle1}
\end{equation}
In the following discussion in this section,  we assume $
0<\epsilon \ll 1$ to be a small parameter. Our objective is to relate 
the acceleration  of the vortex filament $\partial_t {\bf v}$ with  
the  velocity $  {\bf v}$ and local curvature $\kappa$ of the filament.

\begin{figure}
\leftline{\hspace{.0cm} \psfig{figure=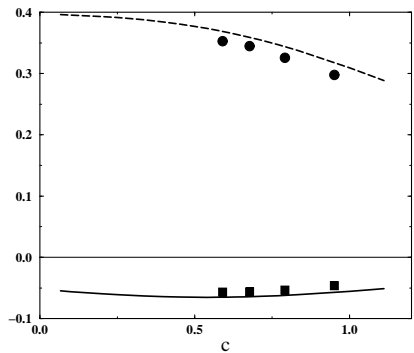,height=3.in}}
\caption{$K_{xx}$ (solid line) and $K_{xy}$ (dashed line)  as  
functions of $c$. Lines are
theoretical results, symbols are the results of two-dimensional  
simulations
from Ref. \protect \cite{akw2} for $\epsilon=0.025$.}
\label{fig1}
\end{figure}

In order to develop the perturbation theory for a weakly-curved
vortex filament in 3D, we begin with
the
stationary  one-armed  isolated spiral
solution to Eqs.\ (\ref{cgle},\ref{cgle1}), which  is of the form
\begin{eqnarray}
A_0(r,\theta) = F(r) \exp i [ \omega t \pm \theta + \psi(r)] \;,   
\label{spir}
\end{eqnarray}
where $(r, \theta)$ are polar coordinates, $\omega = -c -k_0^2 (1-  
\epsilon c) $
is the rotation frequency, and $k_0$ is an asymptotic wavenumber. The  
real functions $F$ and $\psi$
have the following asymptotic behavior $F(r) \to \sqrt{1 - \epsilon  
k_0^2}
\;, \psi^\prime(r) \to k_0 $ for $r \to \infty $ and $F(r) \sim r \;, 
\psi^\prime(r) \sim r$ for $r \to 0 $. The wavenumber $k_0$ of the  
waves emitted by the spiral is determined uniquely
for given $\epsilon,c$ \cite{Hagan}.
For $\epsilon=0$ one has a type of Galilean invariance
and then, in addition to the stationary spiral, there exists
a family of spirals moving with arbitrary constant velocity ${\bf  
v}=(v_x,v_y)$
\cite{akw2},
\begin{equation}
A(r ,t)= F(r^\prime) \exp i[ \omega^\prime t + \theta +  
\psi(r^\prime) +%
\frac{{\bf r}^\prime \cdot {\bf v}}2],  \label{mov}
\end{equation}
where ${\bf r}^\prime = {\bf r} - {\bf v} t$,
$\omega^\prime= \omega -{\bf v}^2/4$,
and the functions $F, \psi $ are those of Eq.\ (\ref{spir}). (This
invariance holds for any stationary solution). For $\epsilon \ne 0$ the 
diffusion term $\sim \varepsilon \Delta A$ destroys the family and
leads to slow acceleration/deceleration
of the spiral proportional to $\epsilon v$, depending
on the value of $\epsilon$.
As  was found in Ref. \cite{akw2}, the equations of motion of the 
spiral core for $\epsilon \ll 1$ assume the form
\begin{equation}
\partial_t {\bf v} +  \epsilon \hat K {\bf v} =0
\label{eom2d}
\end{equation}
with the "friction" tensor
\begin{equation}
\hat K = \left(
\begin{array}{ll}
K_{xx}  & K_{xy} \\
K_{yx} & K_{yy}
\end{array}
\right)
\end{equation}
Because of isotropy the elements of the tensor satisfy the condition:
$K_{xx}=K_{yy}, K_{xy} = - K_{yx} $. Eq. (\ref{eom2d}) can be written in 
a more compact form:
\begin{equation}
\partial_t \hat v + \epsilon \chi \hat v =0, 
\label{solv2d}
\end{equation}
where $\hat v = v_y+i v_x$ is a "complex" velocity and $\chi =  
K_{xx} +i K_{xy} $
is a "complex" friction coefficient.
It was shown in the Ref. \cite{akw2}, that the coefficient
$K_{xx} <0$
for $\epsilon \ll 1$, which implies instability of the spiral core  
(see Fig. \ref{fig1}).
In the limit of the weakly-perturbed  NSE ($c \to \infty$) the friction  
$\chi$
is of the form (see Appendix A):
$\chi \approx -3.45 c^3  \exp[-c \pi ]
( 1 \pm  i 3.92 )$.
Since in general $K_{xy} \ne 0$, the spiral core moves on a logarithmic 
spiral trajectory.

\vspace{-2.0cm}

Let us now consider the dynamics of an almost straight vortex line.
The analysis is conveniently  performed  in a
filament-based coordinate system (see for details \cite{gog}).
The position in space $X$ is represented by
local coordinates $s,\tilde x, \tilde y$, where $s$ is the
arclength along the filament, and ${\bf X} ={\bf R}
(s)+\tilde x {\bf N}(s) + \tilde y {\bf B}(s)$,
where ${\bf R}$ is the coordinate of the filament,
${\bf N}(s)$ the normal and ${\bf B}(s)$ the binormal vector
(see also Fig. 1 in Ref. \cite{gog}).
In this basis the weakly curved filament moving
with velocity ${\bf v}$ can be written in the form
\begin{eqnarray}
A(r ,t)& =& \left(F(r^\prime)+   W(r^\prime,\theta,s)\right) \nonumber \\
  && \exp  \left[ i \left( \omega^\prime t + \theta +\phi_0
+\psi(r^\prime)
  +
 \frac{{\bf r}^\prime \cdot {\bf v}}2
 \right) \right] .
 \label{mov1}
\end{eqnarray}
Here ${\bf r^\prime=r-v} t$, $\phi_0$ is a constant phase, and 
$W$ is the perturbation to the spiral solution which we require to  
be small.
We assume that the velocity vector ${\bf v} $ lies in the plane  
perpendicular
to the vector tangential to the vortex line and may in general  
depend (slowly)
on the arclength $s$.
Substituting $W$ into Eq. (\ref{cgle1}), we obtain at first order in
$\epsilon$ the linear inhomogeneous (assuming $\partial_t v, v , \kappa  
\ll 1$ and
neglecting higher-order corrections $\partial_s v, \partial_s^2 v$ etc.)
equation:
\begin{equation}
\hat L (W, W^*)  = H, 
\label{lin0}
\end{equation}
where
\begin{eqnarray}
\hat L (W, W^*) &=& i \tilde \Delta W+\frac{2 i }{r^2}    
\partial_\theta W+2 i F
\frac{\partial}{\partial r} \frac{ W}{F} \nonumber \\
&& - (1+ i c ) F^2 ( W+W^*)
\label{L}
\end{eqnarray}
and $\tilde \Delta = \partial_r^2 + \frac{1}{r} \partial_r +  
\frac{1}{r^2}
\partial_\theta^2 -\frac{1}{r F } \partial_r ( r \partial_r F) $.
The inhomogeneity $H$ is of the form
\begin{eqnarray}
H &=& \frac{i}{2}   \partial_t {\bf  v } F  - i \partial_t \phi_0 F 
-\epsilon {\bf v} \left( (\nabla F  +i F \nabla ( \theta + \psi )\right)
\nonumber \\
&+&\kappa (\epsilon+i ) \left ( \partial_{ \tilde x}  F+ i F  
\partial_{\tilde
x}  ( \theta + \psi )\right), 
  \end{eqnarray}
where  $(r, \theta)$ are polar coordinates in a local plane spanned on 
the vectors of normal and binormal, and we used the expansion for
the Laplace operator in the local basis
in the limit of
small curvature $\kappa$ and torsion $\tau$:
$\Delta =
 - \kappa \partial_{\tilde x} +
\partial_{\tilde x}^2 +
 \partial_{\tilde y}^2 +\partial_s^2 + ...$.
We use the notation $\hat v =v_B + i v_N$,
where $v_B, v_N $ are the bi-normal and normal
 components of the velocity, respectively (note that
the coordinate $\tilde x$ is directed along the  vector of normal).

Separating real and imaginary part of $W$,
and representing it in the form of a Fourier series
\begin{eqnarray}
{Re W \choose Im W} = \sum_{n=-\infty}^{\infty} {A_n(r) \choose B_n(r)}
 \exp (in\theta), 
\label{ser}
\end{eqnarray}
we arrive at the set of ordinary differential equations for each
azimuthal modes $A_n, B_n$.
Since instability occurs only for the first azimuthal mode,
we consider the equation only for $n=1$ (the equation for $n=-1$ is 
obtained by complex conjugation).
At first order in $\epsilon$ the inhomogeneous linear
equation for the corrections  $A_1, B_1$ is of the form
\begin{eqnarray}
&&\hat{\Delta} A_1 -2 (c F^2 A_1
+ \psi^\prime F \frac{\partial}{\partial r}\frac{B_1 }{F}+ \frac{i  
B_1}{r^2} )
\nonumber \\
&&= -\frac{\epsilon \hat{v}} {2i}  F^\prime + \frac{r \partial_t
\hat{v}}{4i} F
+\frac{ \kappa}{2} F^\prime+\frac{\epsilon  \kappa}{2} (F  
\psi^\prime +\frac{i F}{r})
\nonumber\\
&&\hat{\Delta} B_1 + 2 ( F^2 A_1
+ \psi^\prime F\frac{\partial}{ \partial r}\frac{A_1 }{F}+ \frac{ i  
A_1}{r^2})
\nonumber \\
&& =-\frac{\epsilon \hat{v}} {2i}  (  F \psi^\prime +\frac{i F} r )
-\epsilon \frac{ \kappa}{2} F^\prime+\frac{  \kappa}{2} ( F \psi^\prime
+\frac{i F }{r}), 
\label{lin3}
\end{eqnarray}
where
$ \hat{\Delta}=\partial_r^2+r^{-1} \partial_r -r^{-2} -  
(\partial_r^2 F +1/r
\partial_r F) /F \;$ and
primes denote differentiation with respect to $r$. Equation  
(\ref{lin3}) can be
formally simplified by the following transformation $B_1 = \tilde B_1 - 
\epsilon r  F/4, A_1= \tilde
A_1 $ \cite{gog1}.
After simple algebra,  Eqs. (\ref{lin3})
assume the form (we omit tildes on $A,B$)
\begin{eqnarray}
&&\hat{\Delta} A_1 -2 (c F^2 A_1
+ \psi^\prime F \frac{\partial}{\partial r}\frac{B_1 }{F}+ \frac{i  
B_1}{r^2} )
\nonumber \\
&&= -\frac{\epsilon \bar {v}} {2i}  F^\prime + \frac{r \partial_t
\hat{v}}{4i} F
\nonumber\\
&&\hat{\Delta} B_1 + 2 ( F^2 A_1
+ \psi^\prime F\frac{\partial}{ \partial r}\frac{A_1 }{F}+ \frac{ i  
A_1}{r^2})
\nonumber \\
&& =-\frac{\epsilon \bar {v}} {2i}  (  F \psi^\prime +\frac{i F} r ), 
\label{lin4}
\end{eqnarray}
where $\bar v = \hat v  - i \kappa/\epsilon$.
The solvability condition for Eqs.
(\ref{lin4}) implies an unique  relation between $ \partial_t \hat  
v$ and $\bar v$.
If the solvability condition is fulfilled, the solution $A_1, B_1$
remains finite at $r=0$ and does not grow exponentially for $r \to  
\infty$. Slow,
power-like growth of the solutions is permitted, since the r.h.s. of Eqs
(\ref{lin4}) grows linearly with $r$. Thus the transformation $B_1   
\to  B_1 -
\epsilon r F/4  $ does not change the solvability condition.
Remarkably,  Eqs. (\ref{lin4}) coincide with the equations analysed 
in Ref \cite{akw2} for the case of the acceleration instability of  
a spiral
wave in two dimensions.
The only difference is that the normal  velocity is
modified  by the curvature  $\kappa$. Obviously, it results in the same
solvability conditions as the corresponding equations of the  
two-dimensional case:
\begin{equation}
\partial_t \hat v + \epsilon \chi \bar v =0
\label{solv3d}
\end{equation}
with the same "complex" friction coefficient $\chi$.
The equation of motion (\ref{solv3d}) can be
written in the matrix form
\begin{equation}
{\bf \partial_t v} +  \hat K [\epsilon \bf  v - \kappa N ] =0.
\label{accel2}
\end{equation}
Note, that dropping the
acceleration term in the equation (\ref{accel2}) we recover  the  
result of
Ref. \cite{gog} for $b \to \infty $, since for the ring $v_N  
=\partial_t R$,
$\kappa = -1/R$. Restoring the original scaling $r \to r/\sqrt b$,  
we obtain
$\partial_t R=-b^2/R$. However,  since in three dimensions
the local   velocity
in general varies along the vortex line, even small acceleration
may cause severe  instability of
the vortex line, because the local curvature becomes very large.
Moreover, deviation of the local velocity  from the direction of
normal will lead to stretching and bending of the vortex line.
Thus the acceleration term, which formally can be considered as
a higher order correction to the equation of the motion, may play a  
pivoting
role in the dynamics of a vortex filament.  Our subsequent analysis and 
numerical simulations verify this
hypothesis.

\subsection{Almost straight vortex}

Let us consider
an almost straight vortex parallel to the axis $z$.
We can parameterise the position along the
vortex line  by the $z$ coordinate: $(X_0(z),Y_0(z))$.
Since in this limit
the arclength $s$ is close to $z$,
the curvature correction to the velocity $\kappa {\bf N}$
is simply $\kappa {\bf N}  =(\partial_z^2 X_{0},\partial_z^2 Y_{0})
=\partial_z^2 {\bf r}$, where ${\bf r}= (X_0, Y_0)$.
Using $\partial_t {\bf r} ={\bf v}$, 
from Eq. (\ref{accel2}) we then obtain:
\begin{equation}
\partial_t  {\bf v }+  \hat K [\epsilon   {\bf v} -  \partial_z^2   
{\bf r}] =0, 
\label{accel3}
\end{equation}
Let us now consider  perturbations of the vortex that are periodic  
along $z$.
Due to the linearity of the problem, the solution can be written in the 
form ${\bf  r}  \sim \exp [ i k z + \lambda(k) t]$, where $\lambda  
$ is the
growth rate. We immediately obtain the following relation for  
$\lambda$:
\begin{equation}
 \lambda^2+\chi (
\epsilon \lambda+ k^2)=0. 
\label{lambda}
\end{equation}
Let us consider separately two cases: $k \ll \epsilon $ and $k \gg  
\epsilon$.
For $k \ll  \epsilon$ from Eq. (\ref{lambda}) we derive that $\lambda =-
\epsilon \chi  + O(k^2)$.
For $k \gg \epsilon $ we obtain
$\lambda \approx
  \pm \sqrt {-(K_{xx}\pm i K_{xy } )} k $.
There always exists a root with a large positive real part:   
$\lambda \sim
k \gg \epsilon$.
Therefore, for finite $k$, the growth rate $\lambda(k)$
may significantly exceed the increment of the acceleration  
instability in
2D (corresponding to $k=0$): $ \lambda = -\epsilon \chi  =
 - \epsilon (K_{xy} \pm iK_{xy})$.
We can expect that,
as a result of such an instability, highly-curved vortex filaments
will be formed. Hence, the "small-curvature" approximation  
considered above can be valid
only for finite time. Moreover, one may not expect
this instability to saturate always in a steady-state
configuration with finite curvature (although we will show this  
possibility to exist).
In contrast, we suggest that  frequently reconnection of various parts 
of the filaments, formation of vortex loops etc,  will
result in persistent
spatio-temporal  dynamics of a highly-entangled vortex state.
We expect a fall off of the growth rate $\lambda$ at large $k$  
which is not
captured by the small-curvature approximation used her. This effect  
will be
included in the treatment below.

\section{Numerical results}
In order to verify  our results numerically  and to
follow later development of the instability, 
we performed  simulations
of the three-dimensional CGLE.
We studied a system of $50^3 -  60^3 $ dimensionless units
of Eq. (\ref{cgle1})
with either no-flux or periodic boundary conditions. The numerical  
solution was implemented
on a parallel 16 processors Origin 2000 computer of the 
High Performance 
Computing Center at Argonne National Laboratory.
We applied   an
implicit
Crank-Nicholson algorithm based on iteration scheme
for inversion of  a band matrix. The number of grid points was
$100^3 -  128^3 $. We performed simulations 
in the parameter regime away from amplitude turbulence in
two dimensions
\cite{akw2} for various values of $\epsilon$ and $c$.

We numerically verified our theoretical result for the growth-rate of 
linear perturbation for a straight vortex Eq. (\ref{lambda}).
As an initial condition we selected a straight vortex line with  
small periodic modulation
along the $z$-axis.
We have measured numerically the growth-rate as a function of  
modulation wavenumber
$k$. The results of our simulations are summarized  in Fig. \ref{fig2}.  
As we see from the figure,
the growth rate indeed  increases initially with $k$, and then  
falls off for
large $k$. The theoretical expression (\ref{lambda}) shows reasonable 
agreement with the simulations results for small enough $k$.
The maximum growth rate reached for
intermediate values of $k$ exceeds in this case the growth-rate of the 
two-dimensional acceleration instability ($k=0$) by
more then two orders of magnitude.

\vspace{-2.cm}
\leftline{\hspace{.0cm} \psfig{figure=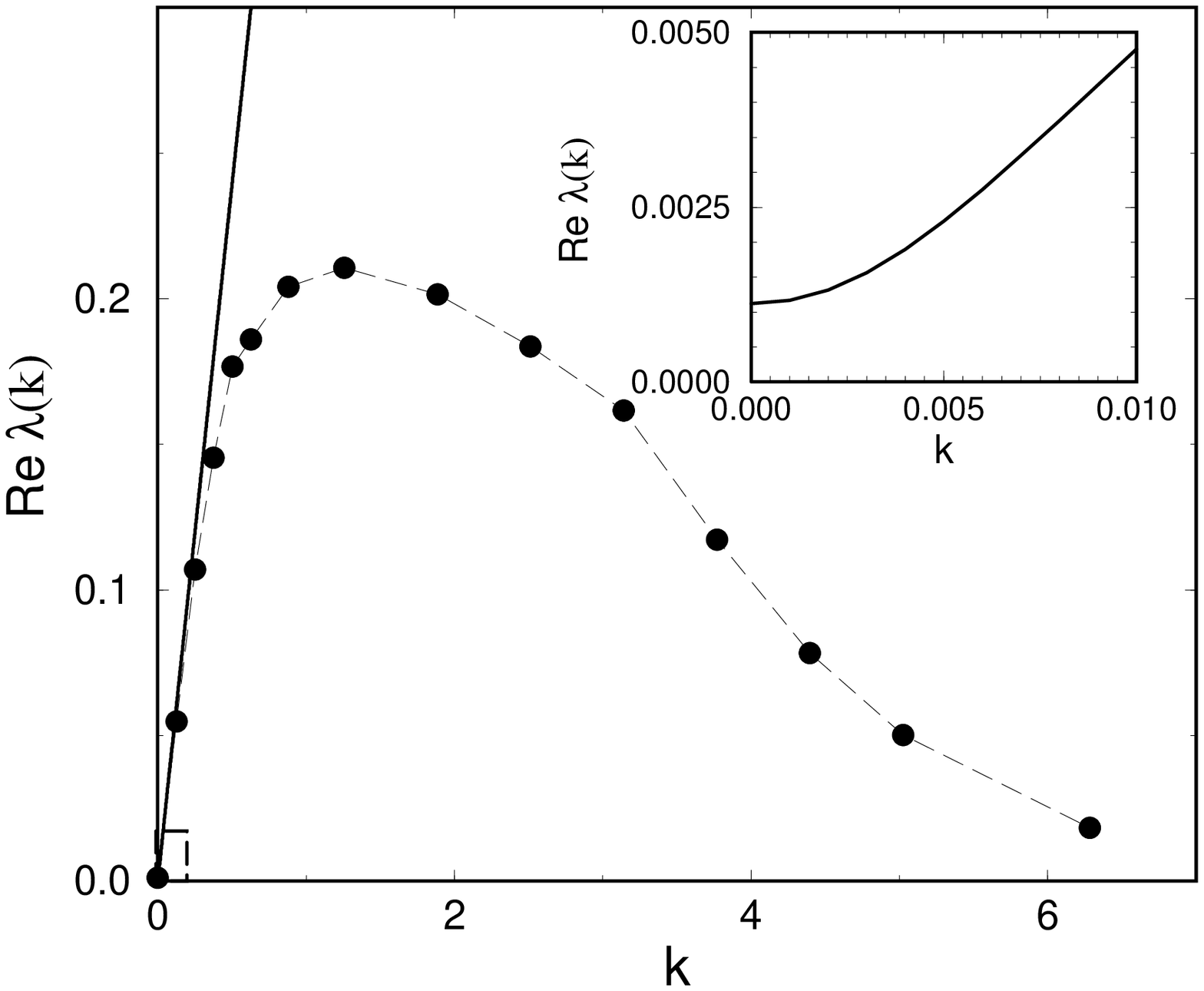,height=3.in}}
\vspace{-.5cm}
\begin{figure}
\caption{
The  growth rate $Re \lambda(k)$ as function of $k$ for
$\epsilon=0.02, c=.1 $. Solid line is the theoretical result
for $k \ll 1$,
dashed line with symbols the result of numerical solution of 3D
CGLE. Inset: blow-up of small $k$ region. }
\label{fig2}
\end{figure}
\leftline{\hspace{.0cm} \psfig{figure=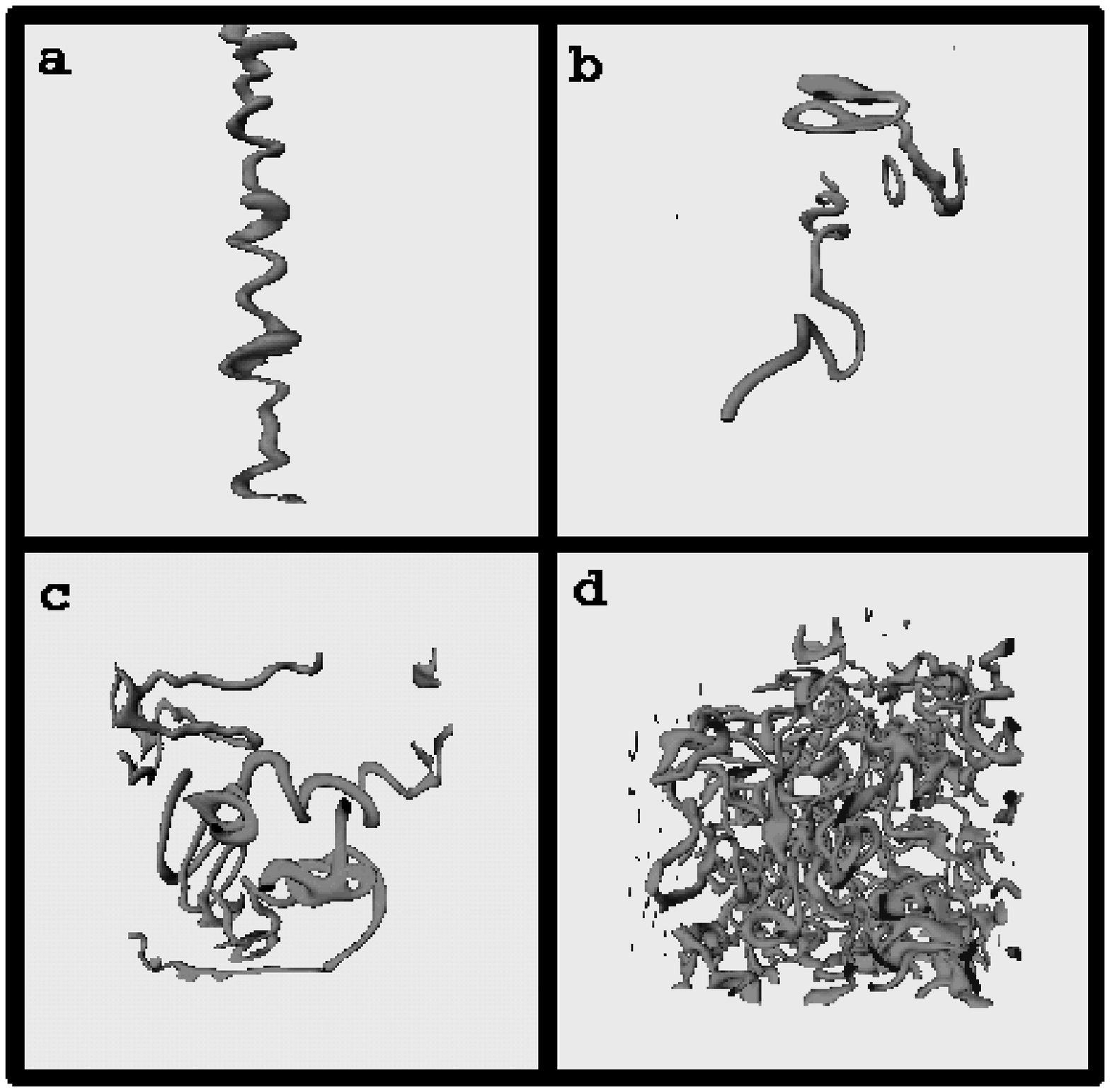,height=3.in}}
\begin{figure}
\caption{
Instability of a straight vortex filament.
3D isosurfaces of $|A(x,y,z)|=0.1$ for
$\epsilon=0.02$, $c=-0.03 $, shown at  four
times: 50 (a), 150 (b), 250 (c), 500(d).
Similar dynamics is observed also for larger value of $\epsilon$.
}
\label{fig3}
\end{figure}

\vspace{-2.5cm}
\leftline{\hspace{.0cm} \psfig{figure=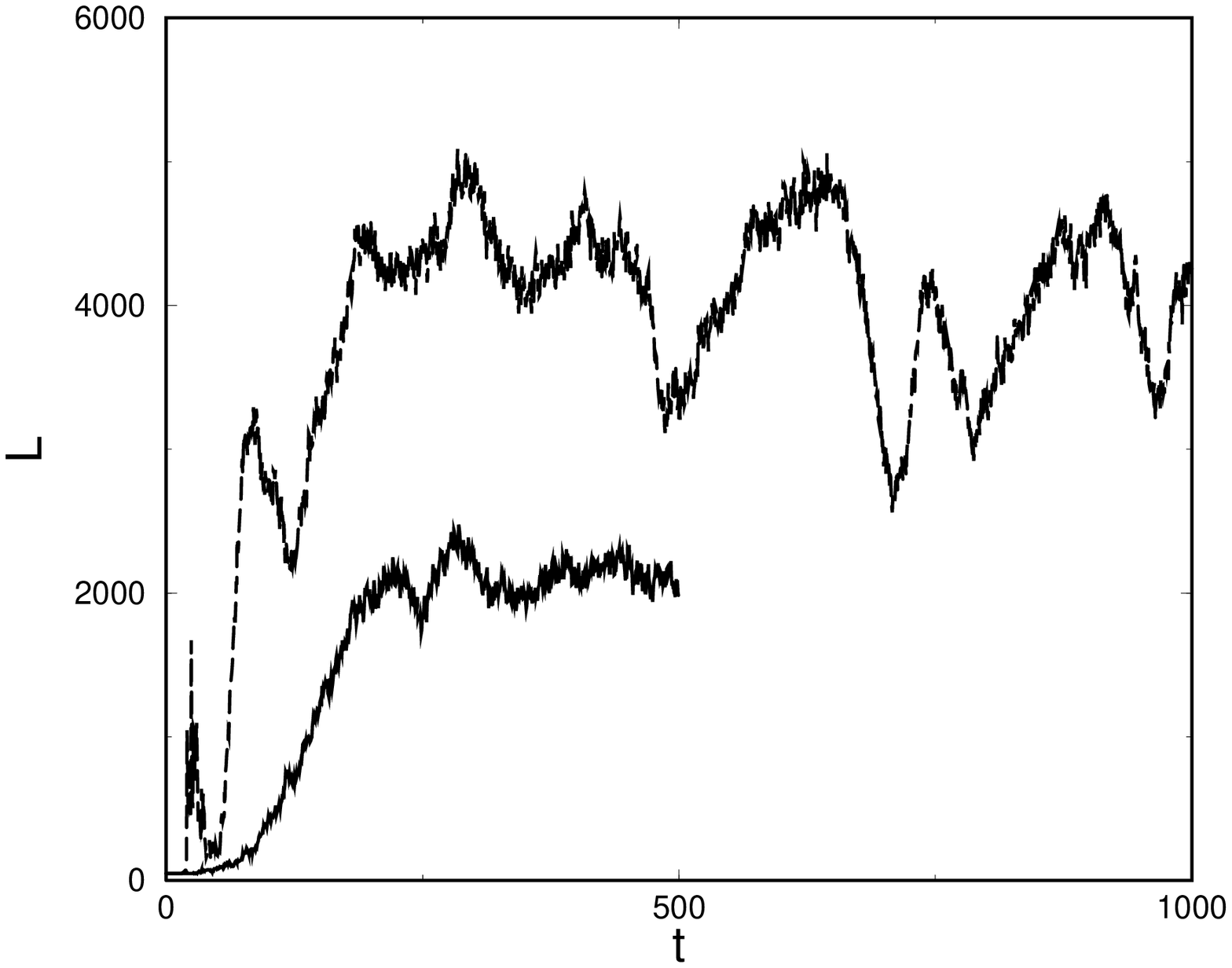,height=3.in}}
\vspace{-.5cm}
\begin{figure}
\caption{
The dependence of filament length $L$ on time.
Solid line corresponds to
$\epsilon=0.02, c=-0.03 $; dashed line corresponds to
$\epsilon=0.02, c=-0.5$.}
\label{fig4}
\end{figure}

The long-time evolution of a straight vortex is shown in Fig. \ref{fig3}.
As initial condition we used a straight vortex perturbed by
a small broad-band  noise. As we can see from the figure,
the length of the vortex line grows. The dynamics seems to
be very rapidly varying in time, and the line intersects itself  
many times
forming numerous vortex loops. The long-time dynamics shows,
however, a saturation when a highly-entangled vortex
state is developed  and the total  length of the line cannot grow  
further due to
a repulsive interaction between closely packed line segments.
The dependence of the line length on time is
shown in Fig. \ref{fig4}.
As a measure of the filament length $L$  we used the following quantity:
 $ L \approx S_0^{-1} \int \Theta(A_0 -
|A(x,y,z)|) dx dy dz$, where
$\Theta $ is the step function: $
\Theta(x)=1$ if $x>0$ and $\Theta=0$ otherwise.
$A_0 = 0.1$ was used as a threshold value to identify the
vortex.
$S_0$ is a constant determined from the condition that for the straight 
line the above integral coincides with the actual length.
We can identify two distinct stages of the dynamics:
first, fast growth of the length;
second, oscillations of the line's length around some
mean value.

\begin{figure}
\leftline{\hspace{.0cm} \psfig{figure=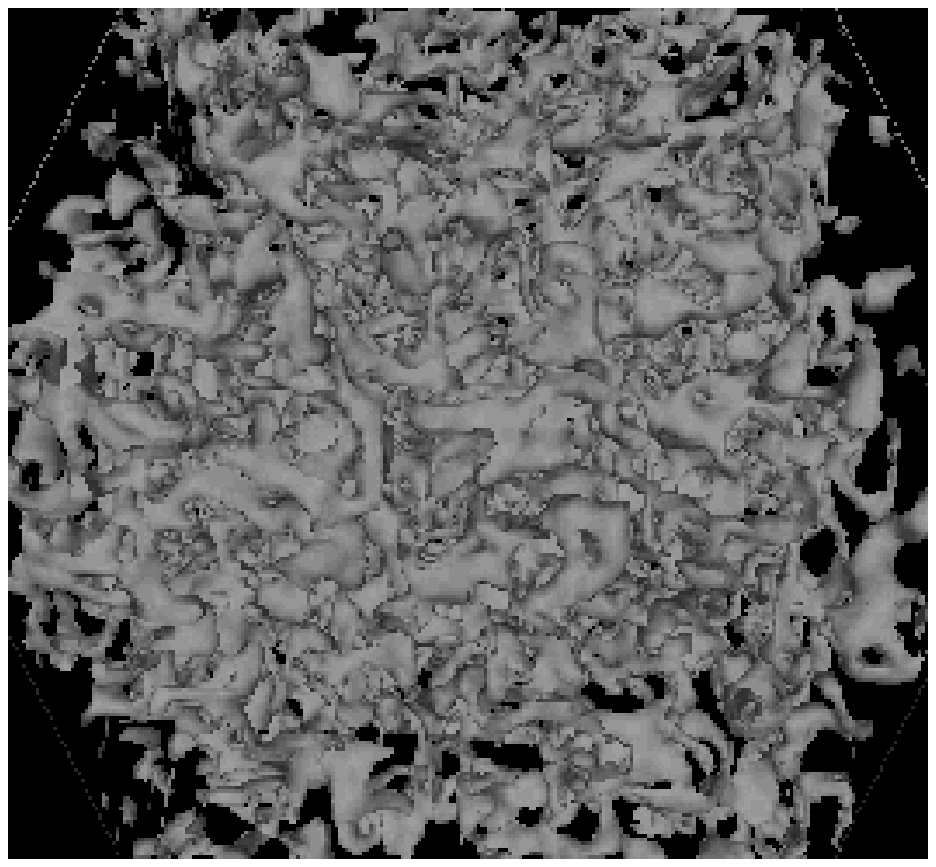,height=1.5in}
\psfig{figure=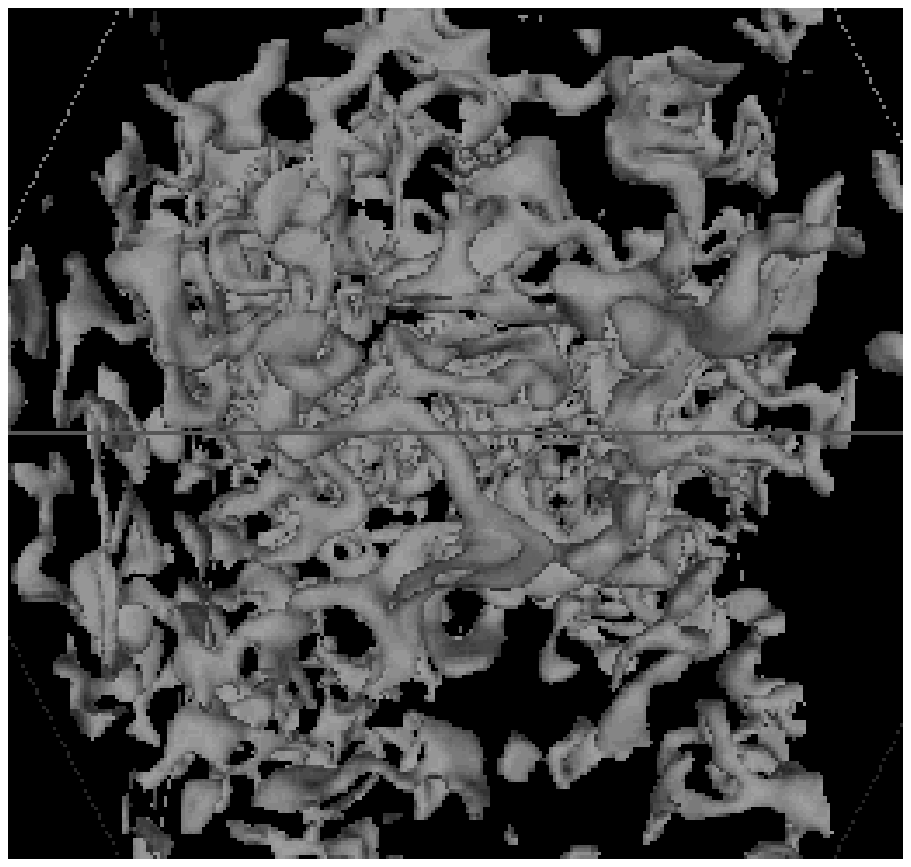,height=1.5in}}
\caption{
Two snapshots of 3D isosurfaces of $|A|$ taken in the regime of
spatio-temporal intermittency,  $\epsilon=0.02, c=-0.5$.
(a) Left image corresponds to  $t\approx 620$ for Fig. \protect  
\ref{fig4}; (b) right image
corresponds to $t\approx 740$.
}
\label{fig4_0}
\end{figure}

Strikingly, for small enough $\epsilon$,  we have observed two  
distinct behaviors
of the total vortex length
depending on the value of $c$. Above a critical value $c_c $
corresponding
approximately to the so-called convective instability range of the  
two-dimensional spiral
($c_c \to 0 $ for $\epsilon \to 0$),
the total length approaches some equilibrium value and does not exhibit
significant fluctuations. On the contrary, for $c< c_c$, the total  
length exhibits large
non-decaying intermittent fluctuations around the mean value.
Figure \ref{fig4} demonstrates the two cases. 
In Fig. \ref{fig4_0} we present snapshots illustrating the  
structure  of
the vortex field in the intermittent case corresponding to the  
maximum of the length
(Fig. \ref{fig4_0}a)
and the minimum (Fig. \ref{fig4_0}b), respectively. One sees, that
in this situation some segments of vortex lines start to expand  
spontaneously,
pushing away other vortex filaments and in such a way making substantial
vortex free holes around them. Then the instability takes over and  
destroys
these almost-straight segments of filament, bringing the system back to a
highly-chaotic state. This dynamics can be considered as a  
three-dimensional
spatio-temporal vortex intermittency, which is an extension of
spiral intermittency discovered in the context of the 
two-dimensional CGLE in
Ref. \cite{akw2}. For even smaller values of $c< -1 $ we observed the
transition to a highly chaotic state, which is an analog of "defect
turbulence" in the two-dimensional CGLE. In this regime small vortex loops
nucleate and annihilate spontaneously, and large vortex filaments play
no role in the dynamics.

\begin{figure}
\leftline{\hspace{.0cm} \psfig{figure=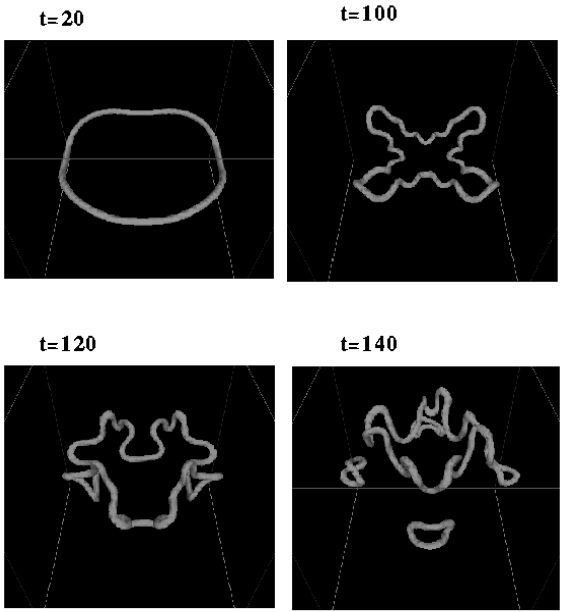,height=3.5in}}
\caption{
Sequence of snapshots  demonstrating the evolution of a vortex
ring for $\epsilon=0.2$ and $c=0.2$. }
\label{fig4_1}
\end{figure}

The evolution of a closed vortex loop is shown in Fig. \ref{fig4_1}. 
Our simulations show that the three-dimensional instability may  
prevent the ring from
collapse, causing the stretching of the loop in the direction  
transversal to
the the  collapse  motion. However, we
have also found that small enough rings
typically collapse, since then the instability described above does  
not have
time to develop substantial distortions of the ring. Even in this
situation the ring exhibits a few oscillations of the radius.
Probably, the oscillations in the collapse  rate in Fig. 2 of Ref.  
\cite{gog} are
caused by the effect of the acceleration term in the equation of  
motion, which
plays an important role even in the stable case.

\section{Limits of three-dimensional instability}
The previous analysis indicates instability of vortex lines in the  
limit
$\epsilon \to 0$ for all $c$. However, it cannot
describe  the boundary  of the instability in the $c,\epsilon$ plane 
since the transition to a stable regime occurs for some finite value of 
$\epsilon$. In order to obtain the stability limit we performed a full
linear stability analysis of a straight vortex solution, which is  
not limited
to small $k$ and $\epsilon$.

The perturbative solution is of the form:
\begin{eqnarray}
A= \left(F(r)+   W(r,\theta,z,t)\right)
 \exp  \left[ i \left( \omega t + \theta +
\psi(r)
 \right) \right] .
 \label{lin2}
\end{eqnarray}
Substituting the ansatz (\ref{lin2}) into the CGLE and performing  
the linearisation
with respect to $W$ as before, then separating the real and  
imaginary part of $W$ and
representing the solution in the form
\begin{eqnarray}
{Re W \choose Im W} = \sum_{n=-\infty}^{\infty} {A_n(r) \choose B_n(r)} 
 \exp \left( i n \theta +  i k z + \lambda(k) t \right), 
\label{ser1}
\end{eqnarray}
we obtain an eigenvalue problem for the eigenvalue $\lambda(k)$.  
Again, we restrict
ourself by the analysis of most dangerous perturbation  harmonics  
with  $n=1$.
The resulting equations are of the from (compare
Eqs. (\ref{lin3},\ref{lin4})):

 \begin{eqnarray}
&&\hat{\Delta} A_1 -k^2 A_1 -2 (c_1 F^2 A_1
+ \psi^\prime F \frac{\partial}{\partial r}\frac{B_1 }{F}+ \frac{i  
B_1}{r^2} )
\nonumber \\
&&= \frac{\lambda(k)}{1+\epsilon^2} ( \epsilon A_1 + B_1)
\nonumber\\
&&\hat{\Delta} B_1 -k^2 B_1 + 2 ( c_2 F^2 A_1
+ \psi^\prime F\frac{\partial}{ \partial r}\frac{A_1 }{F}+ \frac{ i  
A_1}{r^2})
\nonumber \\
&& = \frac{\lambda(k) }{1+\epsilon^2} (\epsilon B_1 - A_1), 
\label{lin5}
\end{eqnarray}
where $c_1 = (\epsilon+c)/(1+\epsilon^2)$ and
$c_2 = (1-\epsilon c)/(1+\epsilon^2)$.
The functions $A_1, B_1$ are subject to the boundary conditions:
 $A_1, B_1$ are bounded at $r=0$ and decay exponentially
for $r \to \infty$.

We solved  Eqs. (\ref{lin5}) in the range of wavenumbers $k$ numerically 
using a matching-shooting method
with Newton iterations from NAG, routine d02agf. Since Eqs.  
(\ref{lin5}) are singular at
$r=0$ and the solution is required on an infinite interval, we  
applied the following
method of solution of this rather difficult eigenvalue problem.
The functions $A_1, B_1$ were expanded in a
series for $r \to 0$ and we used the asymptotic expansion for $r \to  
\infty$.
We replaced the boundary conditions at $r=0$ and $r \to \infty$ by  
new boundary
conditions at sufficiently small $r_0$ and sufficiently large  
$r_e$. The boundary values
were obtained from the corresponding asymptotic expansions, while  
the unknown parameters
of these  expansions
were included in the shooting-matching procedure.
Thus, the numerical
matching procedure was applied on a finite interval $r_0 < r <  
r_e$, where
$r_0$ was typically $10^{-2}$ and $r_e$ was gradually increased until 
the eigenvalue $\lambda$ approaches  its asymptotic value.
Since the unperturbed functions $F, \psi$ are known only numerically, 
we determined them  in the same matching routine, solving
three nonlinear equations (for $F, F^\prime$ and $\psi^\prime$ )
and eight first-order linear equations (the functions $A_1, B_1$ are
complex).  We typically used 5000 mesh points on the interval of
integration.

\begin{figure}
\leftline{\hspace{.0cm} \psfig{figure=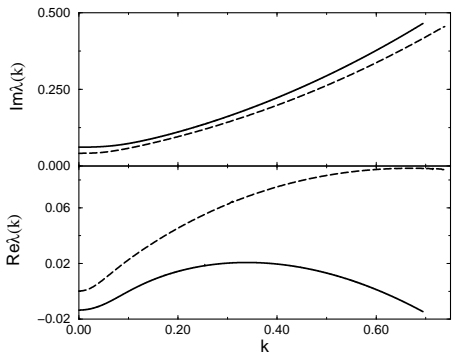,height=3in}}
\caption{
$Re \lambda$ and $Im \lambda$ as functions of $k$ for  
$\epsilon=0.3$ (solid line)
and $\epsilon=0.14$ (dashed line) for $c=0.5$.
}
\label{fig5}
\end{figure}

\begin{figure}
\leftline{\hspace{.0cm} \psfig{figure=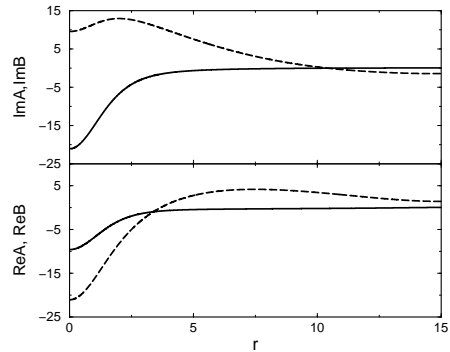,height=3in}}
\caption{
$Re A_1, Im A_1 $ (solid lines) and $Re B_1, Im B_1$ (dashed lines)
as functions of $r$, obtained from numerical solution
of Eqs.( \protect \ref{lin5})  for $\epsilon=0.3, c=0.5$ and $k=0.2$,
corresponding to  $\lambda = 0.0163+i 0.122$.
}
\label{fig5_1}
\end{figure}

The spectrum of $Re \lambda(k) $ for two values of $\epsilon$  
is shown in Fig. \ref{fig5}.
A typical localized core eigenmode, corresponding to this  
three-dimensional
instability is shown in Fig. \ref{fig5_1}.
As we can see from Fig. \ref{fig5},
$Re \lambda(k) $ indeed falls off for large $k$. As expected from  
the previous analysis,
the three-dimensional instability persists beyond the  
two-dimensional core
instability. From  Fig. \ref{fig5} we see that for $\epsilon$  
smaller than about $0.3$ one has
$Re \lambda(0)<0$ (core is stable in two dimensions), however
there is an instability for finite $k$.
The relation between the eigenvalue problem Eq. (\ref{lin5})
and the acceleration instability is presented in Appendix B.

\begin{figure}
\leftline{\hspace{.0cm} \psfig{figure=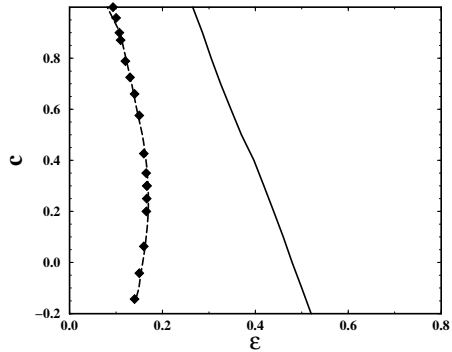,height=3in}}
\caption{
Stability limits in three-dimensions (solid line) and two dimensions
(dashed line), obtained from linear stability analysis. Symbols  
represents
the  limit of the two-dimensional instability, obtained by direct
numerical simulation of the CGLE from the Ref. \protect \cite{akw2}.
Vortex lines and two-dimensional spirals are stable to the left of  
the respective lines.
}
\label{fig6}
\end{figure}

We systematically tracked the boundary for the three-dimensional  
instability
from the condition $\max Re \lambda =0$. The results are shown in  
Fig. \ref{fig6}.
As one can see, the three-dimensional instability occurs in a much wider
parameter range then the two-dimensional core instability. Moreover,
the typical growth-rate in three dimensions is much higher than in  
two-dimensions.

We expect that for $c \to - \infty $ the critical line of
the three-dimensional instability approaches again $\epsilon=0$
line, similarly
to the two-dimensional core instability line \cite{akw2}. However, our 
numerical matching procedure failed to converge for this region of the 
parameters since the critical mode becomes less localized for
$c \to -\infty$.

\vspace{-2.5cm}

\vspace{-2.5cm}

\section{Weakly-nonlinear analysis}
\begin{figure}
\leftline{\hspace{.0cm} \psfig{figure=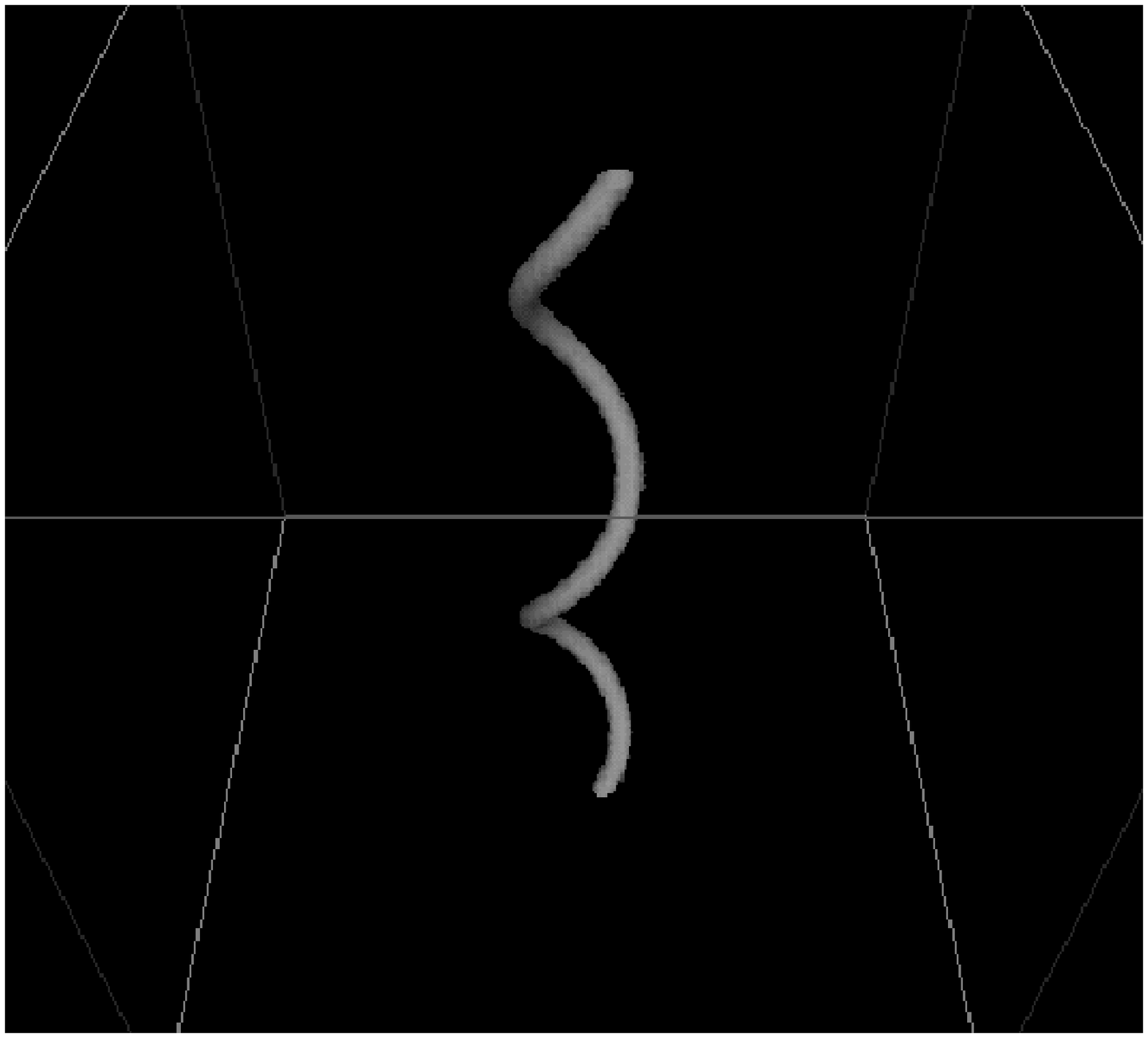,height=1.5in}
\psfig{figure=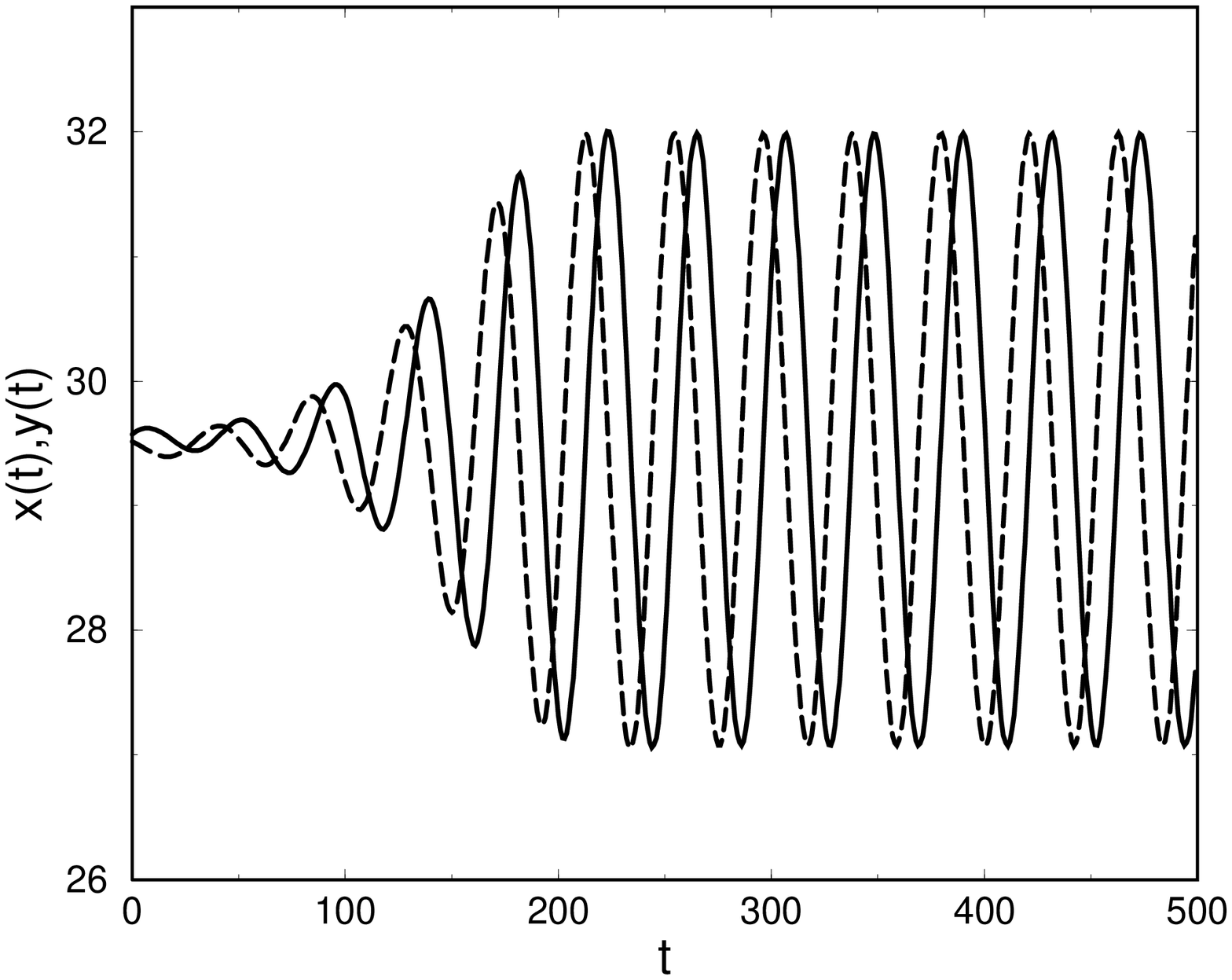,height=2.2in}}
\caption{
(a) Stable travelling helix solution, obtained numerically for
$\epsilon=0.3$ and $c=0.5$.
(b) The coordinates of the vortex line $x(t)$ (solid line) an  
$y(t)$ (dashed line)
as functions of
time for some fixed $z$. A supercritical character of the
bifurcation is apparent.
}
\label{fig7}
\end{figure}

Our numerical simulations near the stability boundary of the  
three-dimensional
instability
have revealed a striking result: the instability saturates, leading to
the formation
of a stable travelling helix solution (see Fig. \ref{fig7}). The  
pitch of the helix is
determined by the most unstable wavelength, and the radius of the helix
vanishes when approaching the stability boundary. The existence of  
the stable
helix solution
can be considered a  result of  saturation of the Hopf bifurcation  
which is thus of
supercritical nature.

\vspace{-2.0cm}

\vspace{-2.0cm}

Although we have never observed saturation of the acceleration  
instability in two
dimensions, here the situation is different, since the most  
unstable mode has
a different spatial structure then the acceleration mode in two  
dimensions.
The helix near the stability limit can be
described in the framework of a weakly-nonlinear analysis for the  
relevant order
parameter, which characterises the local helicity of the vortex line. 

The structure of the amplitude equations can be uncovered from
the following consideration. The growth-rate $\lambda(k)$ is  
symmetric with
respect to $k$: $\lambda(k) = \lambda(-k)$. Therefore, at the  
transition point
two modes with $k_c$ and $-k_c$, corresponding to counter-propagating 
waves of opposite helicity,  simultaneously become unstable. In general, 
$\Omega(k)=Im \lambda(k) \ne 0$. Thus, at lowest order
we shall obtain  two coupled one-dimensional Ginzburg-Landau  
equations for
counter-propagating waves with the amplitudes $U$ and $V$. Close to  
the threshold
the structure of these equations becomes universal:
\begin{eqnarray}
\partial_t U +v_p \partial_z U &=& \lambda(k_c) U + \frac{1}{2}  
\partial_k^2
\lambda(k_c) \partial_z^2U \nonumber \\
&-& \left(a_1 |U|^2 + a_2 | V| ^2 \right) U \nonumber \\
\partial_t V -v_p \partial_z V& =& \lambda(k_c) V + \frac{1}{2}  
\partial_k^2
\lambda(k_c) \partial_z^2 V \nonumber \\
&-& \left(a_1 |V|^2 + a_2 | U| ^2 \right) V
\label{coupl}
\end{eqnarray}
where $v_p = \partial_k \Omega(k_c)$ and $a_1,a_2$ are complex  
constants.
The coefficients of the linear problem can be determined from the
solution of the eigenvalue problem Eqs. (\ref{lin5}). To determine
coupling coefficients $a_{1,2}$ some additional analysis is required. 
In principle, they can be extracted from three-dimensional simulations. 

The system (\ref{coupl}) has been studied intensively in various  
physics contexts
(see e.g. \cite{at,hecke,riecke}). It exhibits diverse   
types of dynamic behaviors,
including  travelling waves, domain walls, sinks, shocks, and  
spatio-temporal
chaos. Thus, in our interpretation, this chaotic behavior corresponds to 
slow nonperiodic
spatio-temporal deformations of the helicoidal structure of the  
vortex line.

Equations (\ref{coupl}) are simplified drastically if one of the
counter-propagating waves
is suppressed, which is the case for  $Re a_1 > Re a_2$. Then this  
system is
reduced to a single one-dimensional complex Ginzburg-Landau equation, 
which is of the form (in a moving frame):
\begin{eqnarray}
\partial_t  U &=& \lambda(k_c) U + \frac{1}{2} \partial_k^2
\lambda \partial_z^2 U  -a_1 |U|^2  U
\label{cgle1d}
\end{eqnarray}
We estimated the parameters of Eq. (\ref{cgle1d}) from our linear  
analysis and
simulations (see Fig. \ref{fig7} b).
We obtained $\lambda(k_c)\approx 0.02095+i 0.1431 $, $ \partial_k^2  
\lambda(k_c)/2 = 0.311-
i 0.5315$ and $a_1 =
a_0^2 (1-i 0.3675)$, where $a_0$ is a  parameter which can be  
scaled out.
For this set of parameters of the one-dimensional  CGLE ($c=-0.3675$ and  
$b=-0.5315/0.311 \approx -1.708$)
the homogeneous solution to Eq. (\ref{cgle1d}) is stable, which  
implies the stability
of a travelling helix solution. However, we can expect that for other  
sets of parameters
of the three-dimensional CGLE the parameters of Eq. (\ref{cgle1d})  
may fall into the
unstable region, e.g the range of amplitude turbulence. This would  
imply chaotic oscillations
of the helix.

We expect that these weakly-nonlinear equations (\ref{coupl}) may  
also serve as
building blocks for the understanding of weak vortex turbulence in the 
CGLE in a certain region of parameters.

\begin{figure}
\leftline{\hspace{.0cm} \psfig{figure=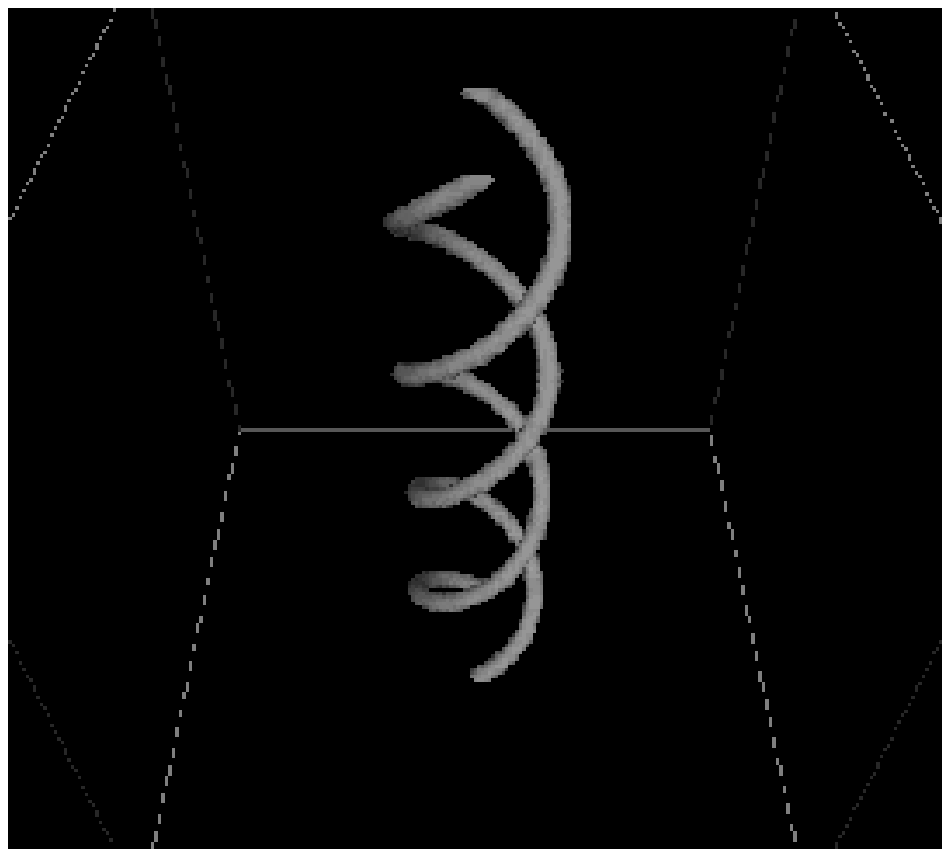,height=1.in}
 \psfig{figure=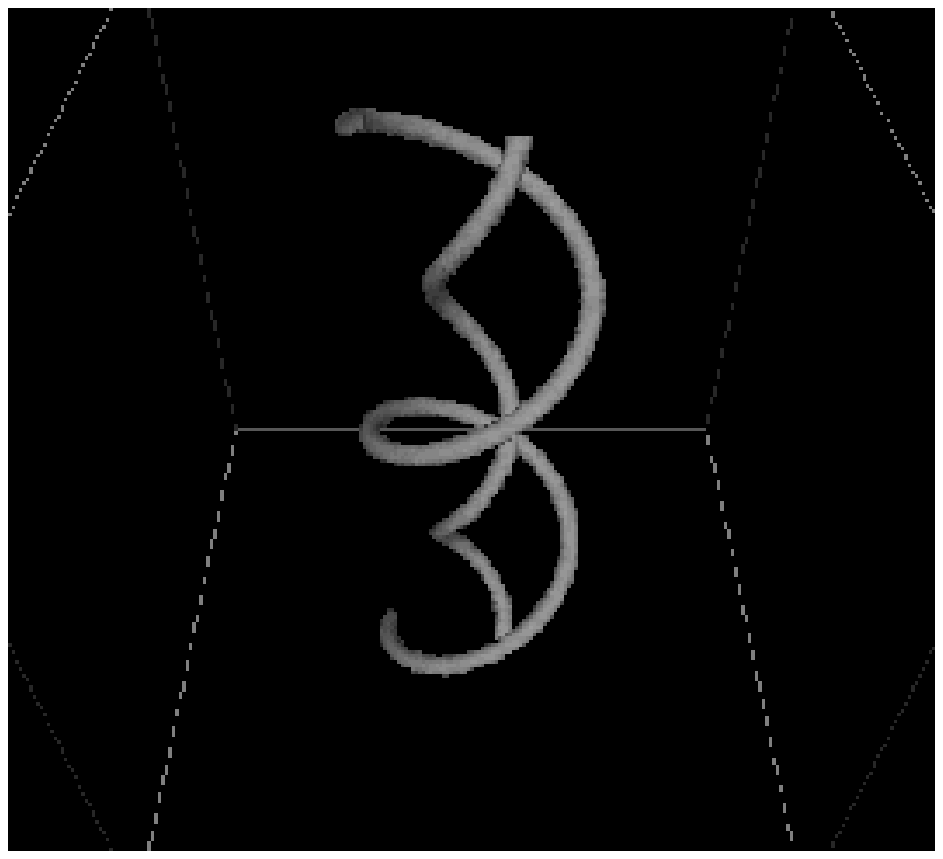,height=1.in}  
\psfig{figure=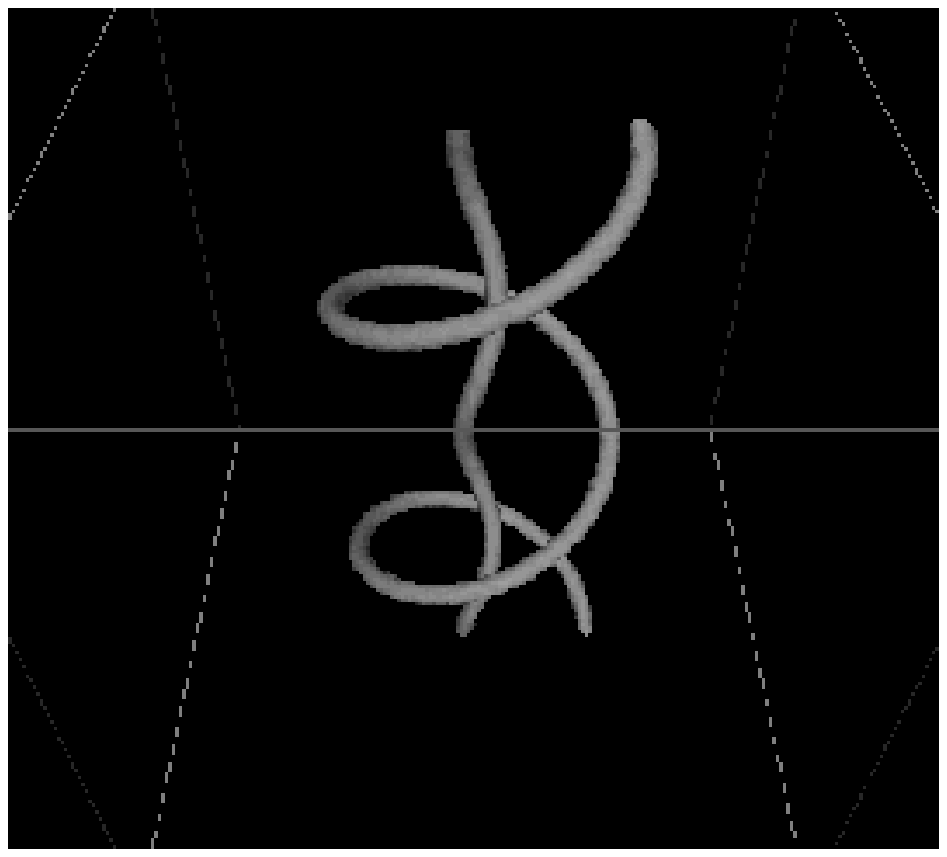,height=1.in}}
\caption{Sequence of snapshots demonstrating the breakdown of
a double helix for $c=0.5 $ and $\epsilon=0.25$. (a) $t=60$,
(b) $t=120$ and (c) $t=240$.
}
\label{fig10}
\end{figure}

An interesting question in this context is: could the helices form  a  
bound state
similarly to spiral waves? One could imagine a stable double-helix state,
similar to the DNA molecule. We have performed a preliminary numerical  
investigation of the
double-helix configuration. The results presented in Fig.  
\ref{fig10} indicate
that the double helix is unstable: the outer helix expands, whereas  
the inner
helix shrinks.
However, we cannot exlude the possibility for existence of a stable 
double helix in some (narrow) parameter range of the CGLE.

\section{Conclusion}
We have derived an equation of motion for the vortex line valid in the 
high-dispersion limit of the CGLE.
Using a general linear stability analysis,  we have found that in a wide 
range of parameters of the CGLE the vortex line is unstable with  
respect to
spontaneous stretching and bending,
resulting in the formation of persistent, dynamic  entangled vortex  
configurations.
In fact vortex lines in three dimensions are unstable in a
much wider range of parameters then two-dimensional spiral waves.
This emphasizes the deficiency of previous approaches relating
local filament velocity only to local curvature.

Slightly beyond the onset of the three-dimensional instability we found  
stable traveling
helix solutions,  which bifurcate supercritically from the straight  
vortex.
Qualitatively similar helicies  have recently been observed
experimentally in heart tissues \cite{pertsov1}.
Note that the acceleration instability in two dimensions is  
subcritical \cite{akw2}.
Unstable helix solutions
can maybe serve as building blocks for a
weakly-nonlinear theory of three-dimensional vortex turbulence.

Let us now discuss implication of our result for the 
well-known phase turbulence 
problem in CGLE. As was found in  Ref. \cite{manneville}, in two dimensions 
phase turbulence is never  a global 
attractor since it is unstable with respect invasion by 
defect turbulence. However, in three dimensions we may expect that 
at least in the region of parameters away from the stretching 
instability,  the vortex rings collapse and  the phase turbulence 
regains its  stability. But inside the three-dimensional 
instability region we may speculate that there is always a possibility 
for creation of a large enough vortex loop which will expand and invade 
the phase turbulence.  

Our result could be verified in experiments with autocatalytic
chemical reactions in gels in the regime of oscillatory instability.
The limit of a large dispersion $b > b_c$  can probably be achieved by 
doping with additional chemicals, thus  changing the relative
mobility of reacting components.

Persistent entangled vortex configurations are known from
numerical simulations of excitable
reaction-diffusion systems \cite{winfree,karma}.
Our preliminary investigation of reaction-diffusion systems shows
that the under-damped core dynamics here is also responsible for  
long-lived
vortex loops and persistent entangled vortex configuration \cite{am}.
In this case, the expansion of the vortex loops is not necessarily  
related to
a "negative line tension" of the filament, but again is the
manifestation of
acceleration effects, similar to the situation in the CGLE.

Recently, the amplitude equation governing the dynamics of an elastic
rod was derived  \cite{tabor}.
The structure of solutions,
found in Ref. \cite{tabor} are remarkably similar to
those of the CGLE.
It is plausible to assume that, in some distinct range
of the parameters,  the equations of motion of the
twisted elastic rode can be reduced to the equations for the vortex  
line in CGLE.
We also speculate that our results are relevant
for inviscid hydrodynamics. In the limit
of $b,c \to \infty$, Eq. (\ref{cgle}) reduces to the defocusing
nonlinear Schr\"odinger
equation (NSE),
which is a paradigm model for compressible inviscid  hydrodynamics. 
Although the vortex lines are stable in the framework of the NSE, the 
corrections arising from the CGLE cause their  destabilisation
and stretching.

\section{Acknowledgements}
We are grateful to A. Newell, D. Levermore, C. Doering and R. Goldstein, 
L. Pismen and W. Zimmermann
for illuminating discussions.
The hospitality of the Max Planck Institute for the Physics of
Complex System at Dresden, where  part of  the work was performed, is 
greatly appreciated.
Most of the computations were performed at the HPCC of  ANL.
This work was supported by the U.S. Department
of Energy under contracts W-31-109-ENG-38 (IA) and
ERW-E420  (AB).
The work of IA was also supported by the NSF,
Office of STC
under contract No. DMR91-20000.

\appendix
\section{Limit of Nonlinear Schr\"odinger Equation}

\subsection{Adjoint mode}
In order to derive the friction coefficient $\chi$ in Eq.
(\ref{solv3d}) we have to fulfill the solvability condition.
In Sec. II this  was done numerically.
The solvability condition  means
the orthogonality of the r.h.s. of Eq. (\ref{lin4}) to the
adjoint mode of  Eq. (\ref{lin4}).
The adjoint equations are of the form:

\begin{eqnarray}
&&\hat{\Delta} A_1^+ -2 \left(c F^2 A_1^+
+ \frac{1}{r F}  \frac{\partial}{\partial r}(r \psi^\prime F B_1 ^+)
- \frac{i B_1^+}{r^2}\right ) \nonumber \\
&&+2 F^2 B_1^+=0
\nonumber \\
&&\hat{\Delta} B_1^+ + 2 \left(
\frac{1}{r F}  \frac{\partial}{\partial r}(r \psi^\prime F A_1^+ )-  
\frac{ i A_1^+}{r^2}
\right)=0. 
\label{lin6}
\end{eqnarray}
The solvability condition of Eqs. (\ref{lin4}) can be expressed in  
terms of
functions $A_1^+,B_1^+$:
\begin{eqnarray}
\partial_t \hat v I_1 -
 2 \epsilon \bar v  I_2 =0, 
\label{ort}
\end{eqnarray}
where
\begin{eqnarray}
I_1&  =&  \int_0^\infty r^2 dr F A^+_1 \nonumber \\
I_2 &=& \int_0^\infty r dr \left (
F^\prime A^+_1 + (F \psi^\prime + i F/r) B^+_1 \right). 
\label{integ}
\end{eqnarray}
From Eq. (\ref{ort}) we readily obtain the friction coefficient
(compare with Eq. (\ref{solv2d}))
\begin{equation}
\chi = -2 I_2/I_1. 
\label{fric}
\end{equation}

For arbitrary $c$ the localized adjoint mode $A_1^+,B_1^+$
can be determined only numerically \cite{akw2}.
However,
for $c \to \infty $, i.e. in the limit of
the perturbed NSE,  the adjoint mode, and $\chi$,  can be calculated  
fully analytically.
For $c \to \infty$ the selected wavenumber $k_0$ vanishes
(see, e.g. \cite{Hagan}),  the solution approaches the
vortex solution of NSE, and Eq. (\ref{lin6}) becomes
self-adjoint. We introduce a small parameter
$\mu=1/c \ll 1$ and assume
$\mu \gg \epsilon$.
Equations (\ref{lin6}) read:
\begin{eqnarray}
\hat{\Delta} A_1^+ &-&2 \left( \frac{F^2}{\mu} A_1^+
+ \frac{1}{r  F}  \frac{\partial}{\partial r}(r \psi^\prime  F B_1 ^+)
- \frac{i B_1^+}{r^2}\right )\nonumber \\
&+&2  F^2 B_1^+=0
\nonumber \\
\hat{\Delta} B_1^+ &+& 2 \left(
\frac{1}{r  F}  \frac{\partial}{\partial r}(r \psi^\prime
F A_1^+ )- \frac{ i A_1^+}{r^2}
\right)=0. 
\label{lin7}
\end{eqnarray}
For $\mu=0$ Eqs. (\ref{lin7}) are self-adjoint and the
localized adjoint eigenmode conincides with the 
complex-conjugated translation mode
\begin{equation}
{ A_1^+ \choose B_1^+ } = {  F^\prime \choose -i  F /r }. 
\label{mod0}
\end{equation}
In order to evaluate Eqs. (\ref{fric}) we take into
account that for any
$\mu \ne 0$ the functions  $A^+_1, B_1^+$ decay
exponentially for $r \to \infty$.
Therefore Eq. (\ref{mod0}) can be considered an approximation valid 
within a finite interval $0<r<R_0$, where the cutoff  $R_0 \gg 1 $ will 
be determined from the matching condition with the outer  
asymptotics of the solution.
For $r \to \infty $ we can simplify Eqs. (\ref{lin7}) using that
$ F^2 \to  1 -\mu/r^2+... $, $\psi^\prime \to k_0 $ and $k_0 \ll 1$:  
\begin{eqnarray}
\hat{\Delta} A_1^+ &-&2 \left(\frac{1} {\mu}  A_1^+
+ \frac{1}{r }  \frac{\partial}{\partial r}(r \psi^\prime  B_1 ^+)
- \frac{i B_1^+}{r^2}\right )
+2  B_1^+=0
\nonumber \\
\hat{\Delta} B_1^+ &+& 2 \left(
\frac{1}{r  }  \frac{\partial}{\partial r}(r\psi^\prime
 A_1^+ )- \frac{ i A_1^+}{r^2}
\right)=0. 
\label{lin8}
\end{eqnarray}
From the first Eq. (\ref{lin8}) we can explicitly express
$A^+_1$ in terms of $B^+_1$, because  for  $r \to \infty$
all  terms in the first Eq. (\ref{lin8}) except $A_1^+/\mu$ 
and $B_1^+$. 
In the first relevant order we obtain
\begin{equation}
A^+_1 = \mu B_1^+
\label{sub1}
\end{equation}
Substituting now Eq. (\ref{sub1}) into the second Eq. (\ref{lin8}), and 
dropping higher-order terms one has
\begin{equation}
\partial_r^2 B_1^+ + \frac{1}{r} \partial_r B_1^+
-\frac{1}{r^2}    B_1^+
+ \frac{2 \mu}{r  }  \frac{\partial}{\partial r}(r\psi^\prime
 B_1^+ ) =0. 
\label{sub2}
\end{equation}
Eq. (\ref{sub2}) is reduced to Bessel's equation by
the substitution
$B_1^+=S \exp(  - \mu \psi) $, leading to
 \begin{equation}
\partial_r^2 S + \frac{1}{r}  \partial_r S
-( \mu^2 k_0^2 +\frac{1}{r^2} )S =0. 
\label{bess}
\end{equation}
It has the localized solution $S= K_1 (\mu k_0 r) $.
Thus, the outer solution is of the form:
\begin{equation}
{ A_1^+ \choose B_1^+ } =  C { \mu  \choose 1  } \exp(-\psi ) K_1  
(\mu k_0 r )
\label{mod1}
\end{equation}
where $C$ is a  constant determined from the matching with
the inner solution Eq. (\ref{mod0}). For $r \to \infty$ the functions decay 
exponentially: $ A_1^+ ,  B_1^+  \sim \exp [- 2 \mu | k_0 | r ]  
/r^{1/2} $,
since $\psi \to k_0 r $. The solutions  (\ref{mod0}) and (\ref{mod1}) 
match in the intermediate region
$r \gg 1$ and $\mu k_0 r \ll 1$. Expanding $K_1 (\mu k_0 r )$ for small
arguments, we find $C= - i \mu k_0$.

\subsection{Friction coefficient}
To calculate the integrals (\ref{integ}) we introduce the cutoff
$R_0 = 1/(\mu k_0) \gg 1$ and split the interval of integration  
into two parts:
$0 < r < R_0$ and $R_0 < r < \infty$. In the first interval we  use the
inner representation of the eigenfunctions Eq. (\ref{mod0}), and in the 
second one the outer representation Eq. (\ref{mod1}).
For the inner interval we obtain
\begin{eqnarray}
I_1^{i} & =&  \int_0^{R_0}  r^2 dr F  F^\prime = b_0 + \mu \log R_0  
\nonumber \\
I_2^{i}  &=& \int_0^{R_0}  r dr \left (
(F^\prime)^2  +  (F/r) ^2 \right) = b_1 +  \log R_0, 
\label{inn}
\end{eqnarray}
where $b_0$ and $b_1$ are some constants.
From the outer integration for  $I_2$ we have
(using $\psi^\prime \to k, F \to 1$ for $r \to \infty$, 
see Eq, (\ref{integ})  ) 
\begin{eqnarray}
I_2^{o}  &\approx& \int_{R_0}^\infty   r (  k_0+i/r)    dr  B^+_1   
\nonumber \\
&\approx & -0.884 -\log( R_0 \mu k_0 ) -i 0.666/\mu. 
\label{outher}
\end{eqnarray}
In order to evaluate the integral $I_1^{o}$ we need the next order of the
function  $A^+_1$ to compensate the logarithmic divergence of
the inner integral. From Eq. (\ref{lin7}) we obtain $A^+_1 = \mu (B^+_1+ i 
B^+_1/r^2) $.
Thus, we have
\begin{eqnarray}
I_1^{o} & =&  \int_{R_0}^\infty r^2 dr [- i \mu k_0 (1 +\frac{i}{r^2} )]
 \mu K_1(\mu k_0 r) \exp(\mu k_0 r) \nonumber \\   &\approx&
 -0.884 \mu  -\mu \log( R_0 \mu k_0)  -  \frac{ 0.4 i}{\mu k_0^2}. 
\label{out1}
\end{eqnarray}
Combining now the outer and inner expansions, we see that $R_0$  
drops out.
The friction coefficient is of the form:
\begin{equation}
\chi = -2
\frac{ -\log(\mu k_0 ) +c_1 -i 0.666/\mu }{ - \mu \log(\mu k_0 )+  
c_0 -i 0.4/(\mu k_0^2)}, 
\end{equation}
where $c_1 = b_1 -0.884, c_0 = b_0 - 0.884 \mu $.
Now, using Hagan's expression for the selected wavenumber  
\cite{Hagan} in
our scaling of the CGLE parameters one finds
$k_0 \approx  2  \mu ^{-3/2} \exp [ - \pi/(2 \mu)-\gamma-0.098] $.
We finally obtain
\begin{eqnarray}
\chi &\approx& -\frac{13.3}{\mu^3 } \exp\left(-\frac{\pi}{\mu}-2  
\gamma -0.196 \right)
\left( 1 + \frac{i\pi}{0.8} \right) \nonumber \\
&& \approx -\frac{3.45}{\mu^3 } \exp\left(-\frac{\pi}{\mu}\right)
( 1 + 3.92 i ).
\label{out3}
\end{eqnarray}
The real part of the friction coefficient
is always negative, which implies the instability of the spiral  
core in two dimensions
and ia stretching instability in three dimensions.

\section{Relation between acceleration instability and  stability  
problem}
The result of Sec. II for weakly-curved vortex filaments
can be formally derived from the linear stability problem Eq.  
(\ref{lin5})
through systematic expansion in $\epsilon$ up to second order.
The eigenvalue problem for $\epsilon \ll \mu$ is  of the form:
 \begin{eqnarray}
&&\hat{\Delta} A_1  -2 (\frac{1}{\mu}  F^2 A_1
+ \psi^\prime F \frac{\partial}{\partial r}\frac{B_1 }{F}+ \frac{i  
B_1}{r^2} )
= (\lambda \epsilon+k^2) A_1 + \lambda B_1
\nonumber\\
&&\hat{\Delta} B_1 + 2 ( F^2 A_1
+ \psi^\prime F\frac{\partial}{ \partial r}\frac{A_1 }{F}+ \frac{ i  
A_1}{r^2})
 = (\lambda  \epsilon+k^2)  B_1 - \lambda A_1 
\label{lin10}
\end{eqnarray}
Now we expand  Eq. (\ref{lin10}) in $\epsilon$.
The solution is represented in the form
\begin{eqnarray}
{A_1 \choose B_1 } &=& {A^{(0)} \choose B^{(0)}} + \epsilon   
{A^{(1)} \choose B^{(1)}}+... \nonumber \\
\lambda & =&  \epsilon \lambda^{(1)} +\epsilon^2 \lambda^{(2)} .... 
\label{expan}
\end{eqnarray}
We consider  $k \sim O(\epsilon)$ and denote $\bar k = k/\epsilon $. 

At zeroth order in $\epsilon$ we simply obtain $\hat L (A^{(0)},  
B^{(0)} ) = 0$,
where $\hat L$ is the r.h.s. of Eqs. (\ref{lin10}). Clearly the  
solution is
the translation mode.
\begin{equation}
 {A^{(0)} \choose B^{(0)}}= {F^\prime \choose F \psi^\prime +  
\frac{i F}{r} }. 
\label{tran}
\end{equation}
At first order in $\epsilon$ we obtain
\begin{equation}
\hat L {A^{(1)} \choose B^{(1)} }= \lambda^{(1)} { F \psi^\prime +  
\frac{i F}{r} \choose -F^\prime}. 
\label{or1}
\end{equation}
The equations have an exact solution corresonding to the 'family mode',
which exists for $\epsilon=0$:
\begin{equation}
 {A^{(1)} \choose B^{(1)} }= -\frac{\lambda^{(1)}}{2} { 0 \choose r F } . 
\label{fam}
\end{equation}
At second order in $\epsilon$ we obtain equations
\begin{equation}
\hat L {A^{(2)} \choose B^{(2)} }={ (\lambda^{(1)} + \bar k ^2)  
F^\prime -\frac{(\lambda^{(1)})^2}{2}
r F \choose   (\lambda^{(1)} + \bar k ^2) ( F \psi^\prime + \frac{i  
F}{r}) } +\lambda^{(2)}.  {B^{(0)} \choose -A^{(0)} }. 
\label{or2}
\end{equation}
These equations have a bounded solution if the solvability
condition is satisfied. It is easy to see that Eqs.
(\ref{or2}) are identical to
Eqs. (\ref{lin4}) if we take into account $\hat v \sim \lambda ,
\partial_t  \hat v = \lambda \hat v  $ and $\kappa = k^2 $. The  
solvability condition implies that
$(\lambda^{(1)})^2 + \chi (\lambda^{(1)}+ k^2 ) = 0 $. The last  
term in Eqs. (\ref{or2})
can be omitted, since it generates a
nonsingular solution (family mode, compare Eq. (\ref{fam})).
Thus, we reproduce the result for a weakly-curved
vortex  Eq.  (\ref{lambda}).


\references{
\bibitem{newell} A.C. Newell, {\it  Envelope Equations},
 American Mathematical Society, Providence, RI, 1974
\bibitem{kuramoto} Y. Kuramoto, {\it Chemical
Oscillations, Waves and Turbulence}, Springer-Verlag, Berlin, 1983
\bibitem{cross} M. Cross and P.C. Hohenberg, \rmp {\bf 65}
851 (1993)
\bibitem{aakw} I.  Aranson, L. Aranson,
L. Kramer and A. Weber, \pra {\bf 46}, R2992 (1992)
\bibitem{akw} I.  Aranson, L. Kramer and A. Weber, \pre
{\bf 47}, 3231 (1993); {\it ibid} {\bf 48}, R9 (1993)
\bibitem{akw2}  I.  Aranson, L. Kramer and A. Weber, \prl
{\bf 72}, 2316 (1994)
\bibitem{chate} G. Huber, P. Alstr{\o}m and T. Bohr, \prl
{\bf 69} 2380 (1992)
H. Chat\^e  and P. Manneville, Physica A
{\bf 224} 348 (1996).
\bibitem{arrechi} F. T.  Arrecchi et al., \prl {\bf 67}, 3749 (1991)
\bibitem{flessel} Q. Ouyang and J.-M.  Flesselles, Nature {\bf 379}, 
6561 (1996)
\bibitem{perez} M. Ruiz-Villarreal, M.  Gomez-Gesteira and V.   
Perez-Villar,
\prl {\bf 78}, 779 (1997)
\bibitem{alt} I. Aranson, H. Levine and L. Tsimring, \prl
{\bf 72}, 2561 (1994)
\bibitem{gold}K.J. Lee, E.C. Cox, and R.E. Goldstein,  \prl
{\bf 76}, 1174 (1996)
\bibitem{siegert} F. Siegert and C.J. Weijer, Proc. Natl. Acad.  
Sci. USA {\bf
89}, 6433 (1992).
\bibitem{gray} R.A. Gray and J. Jalife, Int. J. Bifurcation and  
Chaos {\bf 6},
415 (1996).
\bibitem{vilson} M. Vilson, S. Mironov, S. Mulvey, and
A. Pertsov, Nature {\bf 386}, 477 (1997).
\bibitem{winfree} A. T. Winfree, Physica D {\bf 84}, 126 (1995)
\bibitem{winfree1} A. T. Winfree et al.,  Chaos {\bf 6}, 617 (1996)
\bibitem{bict} V.N. Biktashev,  A. V. Holden, and H. Zhang,
Phil. Trans. R. Soc. London A {\bf 347},  611 (1994).
\bibitem{karma} F. Fenton and A. Karma,
 Scroll Wave Dynamics in a Model of Cardiac Tissue with Restitution 
and Fiber Rotation, to be published, 1997
\bibitem{keener} J. P.  Keener, Physica D {\bf 31}, 269 (1988)
\bibitem{frisch} T. Frisch and S. Rica, Physica D  {\bf 61}, 155 (1992)
\bibitem{gog} M. Gabbay, E. Ott, and P. Guzdar, \prl {\bf 78}, 2012  
(1997).
\bibitem{ab} I.S.  Aranson and A. R. Bishop, \prl {\bf 79}  (1997)
\bibitem{lmn} J. Lega, J.V. Moloney and A.C. Newell, \prl {\bf 73}, 2978
(1994).
\bibitem{Hagan} P. Hagan, SIAM J. App. Math. {\bf 42}, 762 (1982)
\bibitem{gog1} The hint for this substitution can be obtained
from the expansion
of solution $A_0 \exp[ i \delta k_x x ] = A_0 + i \delta k_x x  A_0$, 
found in \protect{ \cite{gog}}, where $A_0$ is unperturbed spiral
solution.
\bibitem{pertsov1} S. Mironov, M. Vinson, S. Mulvey, and
A. Pertsov, J. Phys. Chem.  {\bf 100}, 1975 (1996)
\bibitem{at} I. Aranson and L. Tsimring, \prl {\bf 75}, 3273 (1995). 
\bibitem{hecke} R. Alvarez, M. van Hecke, and W. van Saarloos,
\pre {\bf 56}, R1306 (1997).
\bibitem{riecke} H. Riecke and L. Kramer, preprint, 1997
\bibitem{manneville} P. Manneville and  H. Chat\^e, Physica D, {\bf 96},
30 (1995). 
\bibitem{am} I. Aranson and I. Mitkov, unpublished
\bibitem{tabor} A. Goriely and M. Tabor, \prl {\bf 77}, 3537 (1997);
Physica D {\bf 105}, 20 (1997).
}
\end{document}